\documentclass[fleqn,10pt]{wlscirep}
\usepackage[utf8]{inputenc}
\usepackage[T1]{fontenc}
\usepackage[autostyle=true]{csquotes} 
\usepackage{physics}
\usepackage{comment}

\newcommand{\beq}{\begin{equation}}
\newcommand{\eeq}{\end{equation}}
\newcommand{\beqa}{\begin{eqnarray}}
\newcommand{\eeqa}{\end{eqnarray}}

\newcommand{\cblue}{\color{black}}

\title{Condensate and superfluid fraction of homogeneous Bose gases in a self-consistent Popov approximation}
\author[1]{Cesare Vianello}
\author[1,2,3,4,*]{Luca Salasnich}
\affil[1]{Department of Physics and Astronomy ``Galileo Galilei", University of Padova, Via Marzolo 8, 35131 Padova, Italy}
\affil[2]{Padua QTech Center, University of Padova, Via Gradenigo 6/A, 35131 Padova, Italy}
\affil[3]{National Institute of Nuclear Physics (INFN), Padova Section, Via Marzolo 8, 35131 Padova, Italy} 
\affil[4]{National Institute of Optics (INO) of National Research Council (CNR), Via Nello Carrara 2, 50127 Sesto Fiorentino, Italy}
\affil[*]{luca.salasnich@unipd.it}

\begin{abstract}
  We study the condensate and superfluid fraction of a homogeneous gas of weakly interacting bosons in three spatial dimensions by adopting a self-consistent Popov approximation, comparing this approach with other theoretical schemes. Differently from the superfluid fraction, we find that at finite temperature the condensate fraction is a non-monotonic function of the interaction strength, presenting a global maximum at a characteristic value of the gas parameter, which grows as the temperature increases. This non-monotonic behavior has not yet been observed, but could be tested with the available experimental setups of ultracold bosonic atoms confined in a box potential. We clearly identify the region of parameter space that is of experimental interest to look for this behavior and provide explicit expressions for the relevant observables. Finite size effects are also discussed within a semiclassical approximation.
\end{abstract}

\begin{document}
 
\flushbottom
\maketitle

\thispagestyle{empty}

For a three-dimensional (3D) system of noninteracting bosons in thermodynamic equilibrium, there exists a critical temperature below which a macroscopic fraction of particles occupies the single-particle ground state.  This is the phenomenon of Bose-Einstein condensation (BEC) \cite{einstein, bose}. At zero temperature the gas is completely condensate, and the value of the critical
temperature and the condensate fraction at finite temperature can be computed
exactly.

For an interacting system, simple BEC corresponds to the macroscopic occupation
of one and only one single-particle state. Fragmented BEC, in which more than
one single-particle state is macroscopically occupied, is also possible in
principle \cite{penrose}. Whether or not BEC occurs in a generic bosonic system is a subtle question, and depends strongly on the sign of the effective
inter-particle interaction. Perturbative
treatments (in thermodynamic equilibrium) of the inter-particle interaction starting from the simple BEC state
of the noninteracting gas---like the ones we discuss in this paper---lead to a finite value of the condensate fraction. Given that macroscopic occupation occurs, the fact that it occurs in only one single-particle state is ensured in the dilute limit provided that the effective interaction is
repulsive. In fact in the Hartree-Fock approximation the energy of two identical spinless bosons in different states is greater than that of two such
bosons in the same state, thus macroscopic occupation of more than one state
is energetically unfavorable \cite{leggetts, leggett}. This also means that in a system of weakly repulsive bosons at finite temperature, the tendency to undergo BEC due to quantum statistics should be reinforced by the effect of interactions. Eventually such effect will be overcome by the coherent expulsion from the condensate of most particles by strong interactions (quantum depletion). The condensate fraction should then be a non-monotonic function of the interaction strength.

The question may be addressed using various theoretical approaches, some of which are reviewed by Andersen\cite{andersen}. A non-monotonic behavior of the condensate fraction of a homogeneous gas with zero-range interactions was previously shown by Yukalov and Yukalova\cite{yukalov} working in a self-consistent Hartree-Fock-Bogoliubov approximation \cite{yukalov2, yy2}. Within such frame, the effect is related to the non-monotonic dependence of the anomalous average on the interaction strength. In this paper we employ two different approaches, namely the Bogoliubov theory and a self-consistent Popov approximation. Within these schemes we evaluate numerically and analytically the reinforcing effect of the repulsive interaction towards BEC, showing that, at fixed non-zero temperature, switching on a small interaction causes a sudden increase in the condensate fraction compared to the value for the ideal gas at the same temperature. We also compare it to the superfluid fraction and discuss finite size effects adopting a semiclassical approximation.

Since the first experimental realization of BEC in dilute atomic gases in 1995 \cite{cornell, ketterle}, these have been traditionally produced in harmonic traps. Several methods have therefore been developed to extract uniform-system properties from harmonically trapped ones \cite{nonuniform1, nonuniform2}. For effects such as the quantum depletion, however, only semiquantitative comparison with the theory has been possible, due to complications associated with the inhomogeneity of the clouds or the interpretation of the expansion measurements \cite{nonuniform3}. The realization of BEC in 3D homogeneous potentials in recent years \cite{box-experiment1} has opened the possibility to directly verify theoretical predictions for the homogeneous Bose gas \cite{smith,box-experiment2,box-experiment3}. Hence there is a concrete possibility of testing the predicted non-monotonic behavior in the near future.

We thus consider a homogeneous gas of $N$ spinless bosons of mass $m$ in a 3D volume ${\cal V}$, interacting through a contact potential ${V_\text{int}(\mathbf r'-\mathbf r) = g\delta(\mathbf r'-\mathbf r)}$, where $g=4\pi\hbar^2a_s/m$ is the strength of the two-body interaction with $a_s>0$ the $s$-wave scattering length \cite{andersen, salasnich}. In the functional integral approach the system is described by the Euclidean action $S[\psi^*,\psi] = \int_0^{\hbar\beta}d\tau\int d^3\mathbf x\,\mathcal L(\psi^\star,\psi)$, where $\beta\equiv 1/(k_B T)$, with $T$ the absolute temperature, and
\begin{equation}
    \mathcal L = \psi^*(\mathbf x,\tau)\!\left(\hbar\frac{\partial}{\partial\tau} - \frac{\hbar^2\nabla^2}{2m}-\mu\right)\!\psi(\mathbf x,\tau) + \frac{g}{2}|\psi(\mathbf x,\tau)|^4,
\end{equation}
with $\psi,\,\psi^*$ complex fields satisfying the periodicity condition $\psi(\mathbf x, \hbar \beta) = \psi(\mathbf x, 0)$ and $\mu$ the chemical potential \cite{stoof}. The appearance of BEC is conveniently realized by means of the Bogoliubov prescription \cite{bogoliubov1, bogoliubov2}
\begin{equation}
    \psi(\mathbf x,\tau) = \psi_0 + \psi_1(\mathbf x,\tau),
\end{equation}
where $\psi_0$ is the homogeneous order parameter of the condensate phase transition normalized to the number of condensed atoms, i.e. $|\psi_0|^2 = N_0/\mathcal V \equiv n_0$, and $\psi_1$ describes out-of-condensate fluctuations. At tree level, the value of $n_0$ is fixed by requiring that it sits at the minimum of the part of the action that depends only on $\psi_0,\,\psi_0^*$,
\begin{equation}\label{min}
    n_0 = \frac{\mu}{g},\qquad \mu>0.
\end{equation}

\subsection*{Bogoliubov approximation}
The Bogoliubov theory for the weakly-interacting Bose gas consists of a Gaussian approximation up to second order in the out-of-condensate fluctuations \cite{bogo_orig}. The spectrum of elementary excitations with wavevector of modulus ${k = |\bf k|}$ reads \cite{andersen,salasnich}
\begin{equation}
  \label{specc}
E_k(\mu,n_0) = \sqrt{\left(\varepsilon_k+2g n_0-\mu\right)^2 - (gn_0)^2},\qquad\text{where}\qquad \epsilon_k \equiv \frac{\hbar^2k^2}{2m}.
\end{equation}
The grand canonical potential is \cite{andersen,salasnich}
\beqa
\label{pot}
\frac{\Omega(\mu, n_0, T)}{\mathcal V} &=& \left(\frac{1}{2}g n_0^2
-\mu n_0\right) + \frac{1}{2\mathcal V} \sum_{\mathbf k}
E_k(\mu, n_0) + \frac{1}{\beta\mathcal V}
\sum_{\mathbf k} \ln\left(1-\mathrm e^{-\beta E_k(\mu, n_0)}\right)
\eeqa
and is extremized by
\begin{equation}\label{condv}
    n_0 = \frac{\mu}{g} - \frac{1}{\mathcal V}\sum_{\mathbf k}\frac{2\varepsilon_k + \mu}{E_k(\mu)}\left(\frac{1}{2}+\frac{1}{\mathrm e^{\beta E_k(\mu)}-1}\right),
\end{equation}
where $E_k(\mu) = \sqrt{\varepsilon_k^2 + 2\mu\varepsilon_k}$, which defines the condensate density at the Gaussian (or one-loop) level. From Eq. \eqref{pot} we obtain for the total number density $n$ as a function of the condensate density $n_0$ and $T$,
\beqa\label{n}
  n(n_0,T) = n_0 + \frac{8}{3\sqrt \pi}n_0\left(a_s n_0^{1/3}\right)^{3/2} + \int \frac{d^3\mathbf k}{(2\pi)^3}\frac{\varepsilon_k
    + gn_0}{\left(\mathrm e^{\beta E_k(n_0)}-1\right)
    E_k(n_0)},
\eeqa
where $E_k(n_0) = \sqrt{\varepsilon_k^2 + \left(c_s\hbar k\right)^2}$ and $c_s = \sqrt{gn_0/m}$ is the speed of sound characterizing long-wavelength excitations (see Methods). At zero temperature the last contribution is absent, and approximating $(n_0/n)^{3/2} \simeq 1$ the condensate fraction is given by the Bogoliubov formula \cite{bogo_orig}
\begin{equation}\label{condbog}
    \frac{n_0}{n} \simeq 1- \frac{8}{3\sqrt \pi}\left(a_s n^{1/3}\right)^{3/2},
\end{equation}
which shows that in the interacting case the condensate fraction at $T=0$ is less than one (quantum depletion) and is a monotonically decreasing function of the adimensional gas parameter $\gamma \equiv a_s n^{1/3}$. 

To obtain an expression for the condensate fraction at $T>0$ it is necessary to evaluate iteratively the integral on the right-hand side of Eq. \eqref{n} until obtaining the desired value for the total density. However we can derive two analytical equations that approximate (\ref{n}) in the limits of small temperature or small gas parameter,
\beqa \label{smallt}
      n &\overset{T\to 0}{\simeq}& n_0\left[1+\frac{8}{3\sqrt \pi}\left(a_s n_0^{1/3}
        \right)^{3/2}\right] + \frac{1}{24\sqrt \pi}\frac{t^2}
      {(a_s n_0)^{1/2}}\left(1 + \frac{t^2}{80(a_sn_0)^2}\right)\quad
\eeqa
where $t \equiv m k_B T/\hbar^2$, and
\beqa\label{lowg}
   n &\overset{g\to 0}{\simeq}& n_0\left[1+\frac{8}{3\sqrt \pi}\left(a_s n_0^{1/3}\right)^{3/2}\right] + \text{Li}_{3/2}\left(\mathrm e^{-\beta g n_0}\right)
  \left(\frac{t}{2\pi}\right)^{3/2}
\eeqa
where \text{Li} is the polylogarithm function. For the derivation of these analytical approximations we refer to the Methods. 

At ${g = 4\pi\hbar^2 a_s/m = 0}$, $\text{Li}_{3/2}\left(\mathrm e^{-\beta g n_0}\right) = \text{Li}_{3/2}(1) = \zeta(3/2)$, where $\zeta$ is the Riemann zeta function, and we recover the result for the ideal Bose gas. Eq. \eqref{lowg} thus implies that {\cblue at finite temperature}, for $g\to 0$ the condensate fraction tends to the finite value expected for the ideal gas with positive first derivative (the first derivative is actually infinite at $g=0$, as a property of the polylogarithm function; see Methods, {\cblue Eqs. \eqref{der1}-\eqref{der2}}). Since for large $g$ the condensate fraction is a monotonically decreasing function of $g$, the fact that for very small $g$ it is instead increasing implies that it has a global maximum for a small finite value of the gas parameter. Differently from Yukalov and Yukalova\cite{yukalov}, here the non-monotony is an effect of the dependence of the (normal) thermal depletion on the interaction strength. At a fixed temperature, the quantum depletion increases with the strength of the interaction, starting from zero for zero interaction, while the thermal depletion decreases starting from a finite value for zero interaction. Thus their sum has a global minimum, that translates into a global maximum for the condensate fraction.

\subsection*{Self-consistent Popov approximation}

The preceding discussion is based on the form \eqref{specc} for the spectrum of the elementary excitations, which is obtained in a Gaussian approximation. To evaluate the impact of such approximation we can focus, in particular, on the critical temperature for Bose-Einstein condensation. Setting $n_0 = 0$ in Eq. \eqref{n} we obtain
\beq
n(T_c) = \int \frac{d^3\mathbf k}{(2\pi)^3}\frac{1}{\mathrm
  e^{ \varepsilon_k/k_{\text B} T_c}-1},
\eeq
which gives, for fixed $n$, exactly the value of $T_c$ predicted for the ideal gas. Thus in the Bogoliubov approximation the interaction does not affect $T_c$ with respect to the ideal case. The same holds for the self-consistent Hartree-Fock-Bogoliubov approximation considered in Refs. \cite{yukalov, yukalov2}. However this conclusion is known to be incorrect \cite{andersen}.

The issue is addressed by Kleinert \emph{et al.}\cite{kleinert}, who obtain a self-consistent Popov approximation through a variational method. The idea is to start from the Gaussian approximation \eqref{pot} for the grand canonical potential, expressed as $\Omega = \Omega(\mu, T)$, having used Eq. \eqref{min} to eliminate the dependence from $n_0$, and perform a variational resummation. In order to do so, one introduces an expansion parameter {\cblue $\eta$ (which will eventually be set to one)}, whose power counts the loop order of each term, and a variational parameter $M$, by writing $\mu = M + r\eta$, with $r = (\mu -M)/\eta$. Substituting this into $\Omega(\mu, T)$ and expanding in powers of $\eta$ at {\cblue fixed} $r$, one arrives at an expression for $\Omega^\text{trial}(M, \mu, T)$. According to the principle of minimal sensitivity \cite{stevenson}, one optimizes $\Omega^\text{trial}$ by imposing that $\partial\Omega^\text{trial}/\partial M = 0$ at $M=M^\text{opt}$. The optimized grand potential is then $\Omega(\mu, T) = \Omega^\text{trial}(M^\text{opt}, \mu, T)$. Performing an analogous variational resummation on the condensate density \eqref{condv} and evaluating it for $M = M^\text{opt}$ we obtain
\beqa
  \label{nk}
  n = n_0 + \frac{8}{3\sqrt \pi}\,n\left(a_s n^{1/3}\right)^{3/2} + \int\frac{d^3 \mathbf k}{(2\pi)^3}\frac{\varepsilon_k
      + gn}{\left(\mathrm e^{\beta E_k(n)}-1\right)
      E_k(n)}.
\eeqa
This equation differs from Eq. \eqref{n} by having $n$ in place of $n_0$ in out-of-condensate densities, and amounts to replacing $n_0$ with $n$ in the Bogoliubov spectrum. Hence the considerations on the asymptotic behaviors of (\ref{n}) directly
apply to (\ref{nk}) as well:
\beqa\label{nt0}
n \overset{T\to0}{\simeq} n_0 + \frac{8}{3\sqrt\pi}\,n\left(a_s n^{1/3}\right)^{3/2} +\frac{1}{24\sqrt\pi}\frac{t^2}{(a_s n)^{1/2}}\left(1+\frac{t^2}{80(a_s n)^2}\right)
\eeqa
and
\beqa
 \label{gsmall}
  n \overset{g\to 0}{\simeq} n_0 + \frac{8}{3\sqrt \pi}\, n\left(a_s n^{1/3}\right)^{3/2} +\text{Li}_{3/2}\left(\mathrm e^{-\beta g n}\right)
  \left(\frac{t}{2\pi}\right)^{3/2}.
\eeqa
In particular, also in this approximation the condensate fraction is non-monotonic in the gas parameter. Although the trend is qualitatively similar, numerical evaluation of Eqs. \eqref{n} and \eqref{nk} show that the standard Bogoliubov theory overestimates the condensate fraction compared to the self-consistent Popov approximation, even at zero temperature.

Explicit expressions for the condensate fraction can be obtained
straightforwardly from Eqs. \eqref{nk}, \eqref{nt0} and \eqref{gsmall}. These may be written conveniently in terms of the adimensional gas parameter $\gamma \equiv a_sn^{1/3}$ and the adimensional temperature $T^* \equiv
(mk_\text{B}/\hbar^2 n^{2/3})T$. From Eq. \eqref{nk}, with the change of variable $q \equiv n^{-1/3}k$, we obtain
\begin{equation}\label{nka}
    \frac{n_0}{n} = 1-\frac{8}{3\sqrt \pi}\gamma^{3/2} - I(\gamma, T^*),\qquad\text{with}\qquad I(\gamma, T^*) = \int_0^\infty \frac{dq}{2\pi^2}\,\frac{q^2
    + 8\pi\gamma}{\Bigl(\mathrm e^{(q^2/2T^*)\sqrt{1+16\pi\gamma/q^2}}-1
    \Bigr)\sqrt{1+ 16\pi\gamma/q^2}}.
\end{equation}
Eq. \eqref{nt0} yields
\begin{equation}\label{nt0a}
  \frac{n_0}{n} \overset{T^*\to 0}{\simeq} 1-\frac{8}{3\sqrt \pi}\gamma^{3/2} - \frac{1}{24\sqrt\pi}
  \frac{(T^*)^2}{\gamma^{1/2}}\left(1+ \frac{(T^*)^2}{80\gamma^2}\right)
\end{equation}
and for $T^*=0$ we recover the Bogoliubov formula \eqref{condbog}.
Finally Eq. \eqref{gsmall} gives
\begin{equation}\label{gsmalla}
  \frac{n_0}{n} \overset{\gamma\to 0}{\simeq} 1-\frac{8}{3\sqrt \pi}\gamma^{3/2} - \text{Li}_{3/2}
  \left(\mathrm e^{-\frac{4\pi\gamma}{T^*}}\right)\left(\frac{T^*}{2\pi}\right)^{3/2}.
\end{equation}

\subsection*{Superfluid fraction}

The superfluidity of our system can be characterized following Landau's two-fluid model \cite{landau}. Since the Bogoliubov spectrum is quasi-linear for small momenta, the weakly-interacting Bose gas satisfies Landau's criterion for superfluidity with critical velocity equal to the speed of sound $c_s$. Below the critical velocity, the motion of the fluid through a capillary cannot produce new excitations. However at finite temperature, quasiparticles of thermal excitations are present in the fluid, and the mass flow associated with them is not superfluid, as they can collide with the walls of the capillary and exchange momentum and energy. These quasiparticles account for the normal component of the fluid, and the total density $n$ may be written as $n = n_s+n_n$, where $n_s$ is superfluid density and $n_n$ the normal density.  Collisions establish thermodynamic equilibrium in the gas of excitations, which will be moving with velocity $\mathbf v_n$ with respect to the fluid. Denoting by $E_{\mathbf p}$ the energy of an excitation with momentum $\mathbf p$ in the rest frame of the fluid, the corresponding energy in the quasi-particles' rest frame is $E_{\mathbf p}-\mathbf p \cdot \mathbf v_n$, thus the total momentum density carried by the quasi-particles of thermal excitation is
\begin{equation}
    m n_n \mathbf v_n = \int\frac{d^3\mathbf p}{(2\pi\hbar)^3}\,\mathbf p\,f_\text{B}(E_{\mathbf p}-\mathbf p\cdot\mathbf v_n)
\end{equation}
where $f_\text{B}$ is the Bose distribution. Expanding for small $\mathbf v_n$ and integrating over the directions of $\mathbf p=\hbar\mathbf k$ one obtains the equation for the superfluid fraction,
\begin{align}
    \frac{n_s}{n} &= 1-\frac{\beta}{3mn}\int\frac{d^3\mathbf k}{(2\pi)^3}\frac{\hbar^2k^2\,\mathrm e^{\beta E_k(n)}}{\left(\mathrm e^{\beta E_k(n)}-1\right)^2} = 1-\frac{1}{3T^*}\int_0^\infty\frac{dq}{2\pi^2}\frac{q^4\,\mathrm e^{(q^2/2T^*)\sqrt{1+16\pi\gamma/q^2}}}{\Bigl(\mathrm e^{(q^2/2T^*)\sqrt{1+16\pi\gamma/q^2}}-1\Bigr)^2}
\end{align}
where, in the self-consistent Popov approximation, $E_k(n)$ is the Bogoliubov spectrum with $n$ replacing $n_0$. In the limit of small $T^*$ this is approximated by
\begin{equation}
    \frac{n_s}{n} \overset{T^* \to 0}{\simeq} 1-\frac{1}{720\sqrt\pi}\frac{(T^*)^4}{\gamma^{5/2}}.
\end{equation}

\subsection*{Finite size effects}

The non-monotonic behavior of the condensate fraction is even more pronounced if we include finite-size effects. Within a semiclassical approximation, these are taken into account by introducing an infrared (IR) cutoff $k_\text{IR}$, whose inverse is related to the system size \cite{lanaro}. Let us suppose that the introduction of the cutoff does not break the spherical symmetry in momentum space, that is the case for a spherical confinement. The regularized zero-temperature density, i.e. the second term of Eq. \eqref{n}, then becomes
\begin{align}\label{ngfs}
    n_{g,\,\text{fs}}^{(0)} &= \frac{1}{2}\int_{k_\text{IR}}\frac{d^3\mathbf k}{(2\pi)^3}\frac{\varepsilon_k + gn}{E_k(n)} = \frac{1}{12\pi^2}\left(8\pi a_s n - k_\text{IR}^2\right)\sqrt{16\pi a_s n + k_\text{IR}^2}
\end{align}
The quantity on the right-hand side of Eq. \eqref{ngfs} is positive definite only for $k_\text{IR} < \sqrt{8\pi a_s n}$, which implies a bound on the system size $L$. Taking $k_\text{IR} \sim 1/L$ and $n \sim N/L^3$, one gets $L \lesssim 8\pi a_s N$. We can physically interpret this condition in terms of the Bogoliubov spectrum, for if $k_\text{IR}$ is too large we are effectively cutting off the quasi-linear interval of the spectrum that corresponds to the condensate phase. 

With the change of variable $q_\text{IR} \equiv n^{-1/3}k_\text{IR}$, the condensate fraction thus reads
\begin{align}
    \left(\frac{n_{0}}{n}\right)_\text{fs} &= 1 - \frac{1}{12\pi^2}\left(8\pi\gamma-q_\text{IR}^2\right)\sqrt{16\pi\gamma + q_\text{IR}^2} - I(\gamma, T^*; q_\text{IR})
\end{align}
where $I(\gamma, T^*; q_\text{IR})$ corresponds to the quantity defined in Eq. \eqref{nka} with the integration restricted to the interval $[q_\text{IR}, +\infty)$. The value of $q_\text{IR}$ is bounded by $q_\text{IR} < \sqrt{8\pi\gamma}$.
Similarly, the superfluid fraction is
\begin{equation}
    \left(\frac{n_s}{n}\right)_\text{fs} = 1-\frac{1}{3T^*}\int_{q_\text{IR}}^\infty\frac{dq}{2\pi^2}\frac{q^4\,\mathrm e^{(q^2/2T^*)\sqrt{1+16\pi\gamma/q^2}}}{\Bigl(\mathrm e^{(q^2/2T^*)\sqrt{1+16\pi\gamma/q^2}}-1\Bigr)^2}.
\end{equation}
Notice that at zero temperature the superfluid fraction is always one, and is not influenced by the finite size.

If the confining potential is not spherical, the IR cutoff should be introduced accordingly. For instance, in the case of a cubic box of length $L$ and assuming vanishing boundary conditions, we would introduce $k_\text{IR}=\pi/L$ and, changing to adimensional variables, perform the integrations as
\begin{align}\label{fmod}
    \left(\frac{n_0}{n}\right)_\text{fs} &= 1 - \biggl(\frac{8}{3\sqrt \pi}\gamma^{3/2} - \int_U \frac{d^3\mathbf q}{(2\pi)^3}\frac{q^2 + 8\pi\gamma}{q^2\sqrt{1- 16\pi\gamma/q^2}}\biggr) - 8 \int_D\frac{d^3\mathbf q}{(2\pi)^3}\frac{q^2 + 8\pi\gamma}{\Bigr(\mathrm e^{(q^2/2T^*)\sqrt{1+16\pi\gamma/q^2}}-1\Bigl)q^2\sqrt{1-16\pi\gamma/q^2}}
\end{align}
with $U = [-q_\text{IR}, q_\text{IR}]\times[-q_\text{IR}, q_\text{IR}]\times[-q_\text{IR}, q_\text{IR}]$ and $D = [q_\text{IR}, +\infty)\times[q_\text{IR}, +\infty)\times[q_\text{IR}, +\infty)$, and
\begin{equation}
    \left(\frac{n_s}{n}\right)_\text{fs} = 1-\frac{8}{3T^*}\int_D\frac{d^3\mathbf q}{(2\pi)^3}\frac{q^2\,\mathrm e^{(q^2/2T^*)\sqrt{1+16\pi\gamma/q^2}}}{\Bigl(\mathrm e^{(q^2/2T^*)\sqrt{1+16\pi\gamma/q^2}}-1\Bigr)^2}.
\end{equation}
Requiring the term between round brackets in Eq. \eqref{fmod} to be positive definite will impose an upper bound on $q_\text{IR}$, and thus $k_\text{IR}$, which implies a bound on the system size.

\section{Results}

\begin{figure}[t!]
\centering
\includegraphics[width=0.48\textwidth]{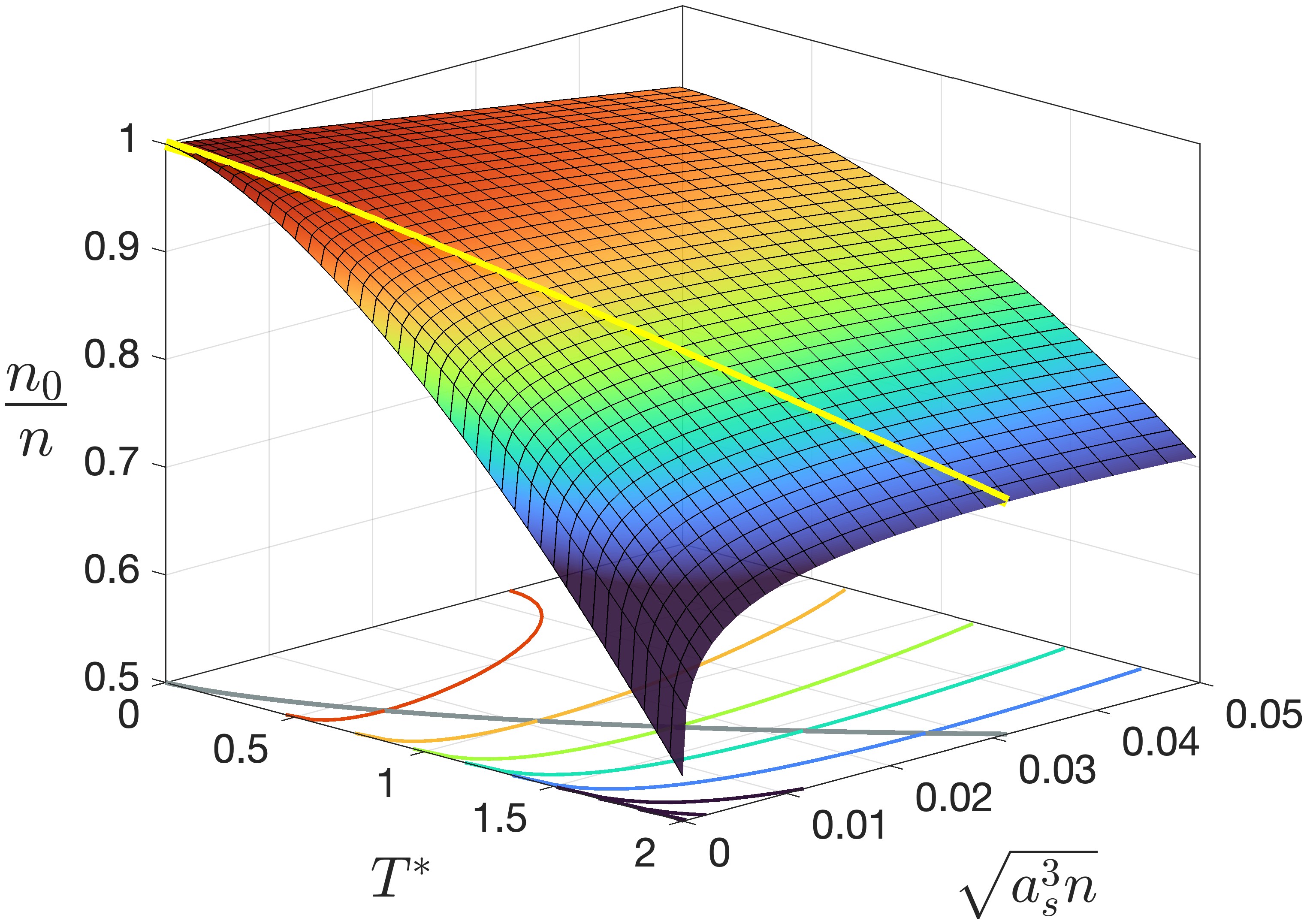}
\caption{Condensate fraction as a function of $\gamma^{3/2}=\sqrt{a_s^3 n}$ and the adimensional temperature $T^* = (mk_\text{B}/\hbar^2 n^{2/3})T$. The condensate fraction at $T^*=0$ decreases linearly with $\sqrt{a_s^3 n}$ according to the Bogoliubov formula \eqref{condbog}. The yellow line follows the maximum of $n_0/n$ as $T^*$ varies.}
\label{condxi-fig}
\end{figure}

\begin{figure}[t!]
\centering
\includegraphics[width=0.48\textwidth]{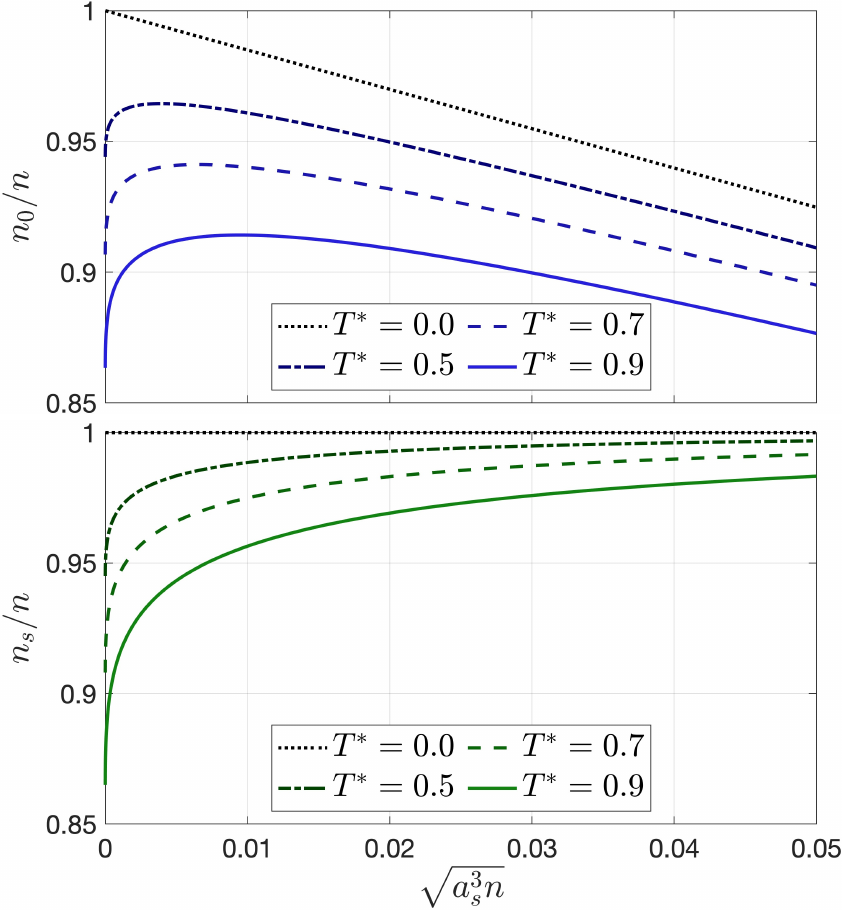}
\caption{Comparison between the condensate fraction $n_0/n$ (upper panel) and the superfluid fraction $n_s/n$ (lower panel) as functions of $\sqrt{a_s^3 n}$, for $T^* = 0.0,\,0.5,\,0.7,\,0.9$.}
\label{cond-super}
\end{figure}

Fig. \ref{condxi-fig} shows the condensate fraction obtained from Eq. \eqref{nka} as a function $\gamma^{3/2} = \sqrt{a_s^3 n}$ and $T^*$, while in the upper panel of Fig. \ref{cond-super} we plot the condensate fraction as a function of $\gamma^{3/2}$ for fixed $T^*$, showing clearly the non-monotonic behavior at finite temperature. Numerical evaluation of Eq. \eqref{nka} leads to the following results:\\
(i) The value of $\gamma$ corresponding to the maximum of $n_0/n$, which we denote $\gamma_\text{max}$, increases linearly with $T^*$:
\begin{equation}\label{gammamax}
    \gamma_\text{max} = aT^*,\qquad a = 0.0498^{+1}_{-1}.
\end{equation}
(ii) The maximum of $n_0/n$, which we denote $(n_0/n)_\text{max}$, decreases with the power $3/2$ of $T^*$:
\begin{equation}
    \left(\frac{n_0}{n}\right)_\text{max} = 1-b(T^*)^{3/2},\qquad b = 0.1005^{+1}_{-1}.
\end{equation}
(iii) As the gas parameter increases, $n_0/n$ decreases more slowly
with $T^*$, leaving a larger residual condensate fraction as
$T \to T^{(0)}_c$. Consequently the critical BEC temperature $T_c$ increases with the gas parameter. Solving numerically Eq. (\ref{nka}) with $n_0 = 0$ gives that the shift of $T_c$ is proportional
to the square root of the gas parameter. In fact, this can also be proven
analytically \cite{kleinert}:
\begin{equation}
  \frac{\Delta T_c}{T_c^{(0)}}=\frac{4\sqrt{2\pi}}{3 \zeta(3/2)^{2/3}}
  \sqrt{\gamma}
  \simeq 1.762\sqrt{\gamma}.
\end{equation}
This results contrasts with Monte-Carlo simulations \cite{svistunov, arnold}
and precise high-temperature calculations \cite{kleinert2, kastening}, which
instead show that the shift $\Delta T_c/T_c^{(0)}$ is linear in the gas
parameter, with a slope $C_0\simeq 1.3$. This is due to the fact that the
self-consistent Popov approximation ignores the pile-up of infrared
singularities to high orders at the critical point \cite{kleinert}, and
therefore it can only be considered reliable far from the critical point. In general this is a problem common to any perturbative approach, since the physics at the phase transition is inherently non-perturbative.

These results should be compared with the ones reported by Yukalov and Yukalova\cite{yukalov} (see in particular Fig. 3 therein, taking into account that on the $x$ axis there is $\gamma$, while we have $\gamma^{3/2}$). The behavior is qualitatively similar, although in Yukalov and Yukalova it seems more pronounced. Indeed, Yukalov and Yukalova consider a much wider range of $\gamma$, between 0 and 1, while we limit ourselves to $\gamma \lesssim 0.15$. The temperature range considered is also wider, reaching up to the critical value where the condensate fraction becomes zero. The reason why our results are limited to the indicated values of $\gamma$ will be addressed in the Discussion. Concerning the results of Yukalov and Yukalova, we point out that their validity for large values of $T$ and $\gamma$ is questionable. As already mentioned, the critical temperature is predicted to be the same as that for the ideal gas, which is known to be incorrect. Furthermore, Monte-Carlo calculations at $T=0$ show that, while the approach of Yukalov \emph{et al.}\cite{yukalov, yukalov2, yy2} works well for $\gamma \lesssim 0.45$, it deviates strongly for larger values of $\gamma$ \cite{rossi}. In the limit $\gamma \to 0$, such approach reproduces the standard Bogoliubov theory \cite{yukalov2}. Therefore the self-consistent Popov approximation provides an improved description of the system in the small $\gamma$ region. This is the region of experimental interest for testing the non-monotonic behavior using the current setups of ultracold Bose gases confined in a box potential.

Lopes \emph{et al}.\cite{smith} studied a homogeneous gas of $^{39}$K atoms ($m\simeq 6.48\times 10^{-23}$ g) of density $n\simeq 3.5\times 10^{11}$ cm$^{-3}$. By varying the interaction parameter in the range $0.004 \lesssim \sqrt{a_s^3 n} \lesssim 0.03$ the authors obtained data consistent with the Bogoliubov theory of quantum depletion for a temperature between 3.5 and 5 nK, corresponding to $0.57 \lesssim T^* \lesssim 0.81$, observing the decreasing trend of $n_0/n$ expected for $\sqrt{a_s^3 n}>\gamma_\text{max}^{3/2}$. However the lowest value of $\sqrt{a_s^3 n}$ achieved in this setup is still too large to clearly test the existence of an interval where the condensate fraction is increasing. Indeed Eq. \eqref{gammamax} gives $\gamma_\text{max}^{3/2}(T^*\simeq0.57) \simeq 0.005$ and $\gamma_\text{max}^{3/2}(T^*\simeq0.81) \simeq 0.008$, which are quite close to the first experimental point at $\sqrt{a_s^3 n}\simeq 0.004$, whereas to test the region where the condensate fraction is increasing would require to reach values of $\sqrt{a_s^3 n}$ much smaller than $\gamma_\text{max}^{3/2}$.

Although the approximations we have discussed are only valid at low temperature \cite{andersen}, we expect the non-monotonic dependence of the condensate fraction on the interaction strength to be a characteristic of weakly interacting Bose gases at any finite temperature below $T_c$. This behavior must in fact be present based only on the well-established facts that the interaction causes (i) the quantum depletion of the condensate at zero temperature and (ii) a positive shift of the critical temperature $T_c$.

In Fig. \ref{cond-super} we compare the condensate and superfluid fractions as functions of $\sqrt{a_s^3 n}$ for fixed $T^*$. The two coincide in the case of zero interaction. In the interacting case, instead, the condensate fraction behaves quite differently from the superfluid fraction; finite temperature depletes both of them, however the dependence on the interaction strength is noticeably different. At ${T^* = 0}$ the system is completely superfluid, independently of the strength of the interaction, and for finite $T^*$ the superfluid fraction is a monotonically increasing function of the gas parameter.

\begin{figure}[t!]
\centering
\includegraphics[width=0.48\textwidth]{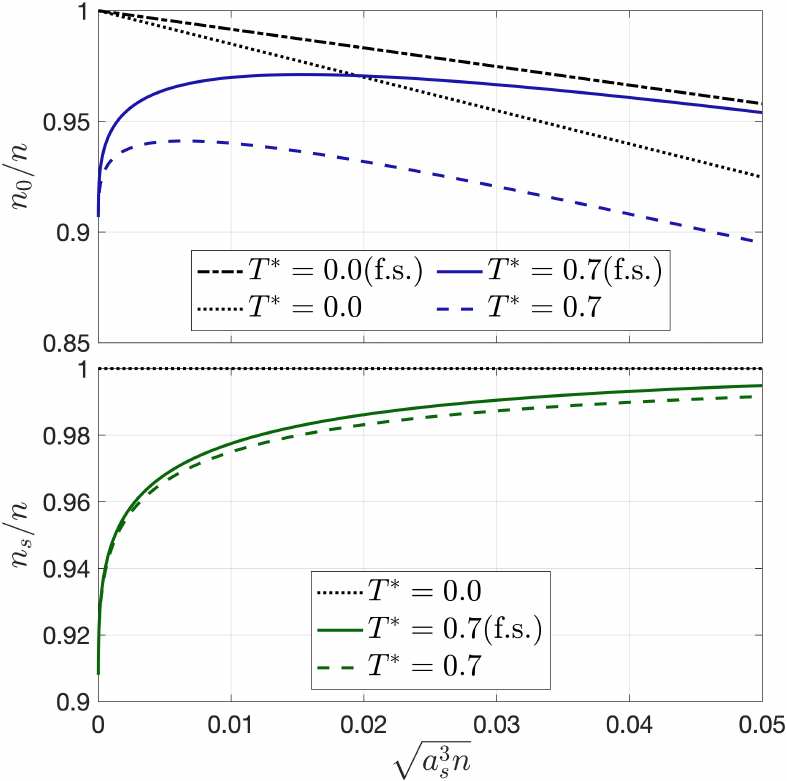}
\caption{Comparison between the condensate fraction $n_0/n$ (upper panel) and the superfluid fraction $n_s/n$ (lower panel) in the thermodynamic limit and at finite size (f.s.), as functions of $\sqrt{a_s^3 n}$, for $T^* = 0.0,\,0.7$. The IR cutoff is fixed at $q_\text{IR} = \sqrt{4\pi\gamma} = \sqrt{4\pi a_s n^{1/3}}$.}
\label{finitesize}
\end{figure}

In Fig. \ref{finitesize} we compare the finite size results at fixed $T^*$, for $q_\text{IR}=\sqrt{4\pi\gamma}$, with the corresponding results in the thermodynamic limit, where $q_\text{IR}\to0$. (The chosen value of $q_\text{IR}$ is such that the bound $q_\text{IR}<\sqrt{8\pi\gamma}$ is satisfied and the the difference with the case $q_\text{IR} = 0$ is clearly visible). The introduction of the IR cutoff effectively reduces both the quantum and the thermal depletions, leading to larger condensate and superfluid fractions compared to the case $q_\text{IR}=0$, also for zero interaction.  Although quantitatively different, the results in the case of a cubic confining potential are qualitatively analogous to those shown in Fig. \ref{finitesize}.

\section{Discussion}

The approximations we have adopted are only valid for small values of the gas parameter. This limitation can be deduced from a purely thermodynamic argument. The thermodynamics of the system is characterized by the grand canonical potential $\Omega(\mu,\,T)$, obtained by imposing to \eqref{pot} the condition \eqref{min}. At zero temperature, it reads\cite{andersen, salasnich}
\begin{equation}
    \frac{\Omega(\mu)}{\mathcal V} = -\frac{\mu^2}{2g} + \frac{8}{15\pi^2}\left(\frac{m}{\hbar^2}\right)^{3/2}\mu^{5/2},\qquad \mu>0.
\end{equation}
where the second term is the well-known Lee-Huang-Yang correction\cite{lhy}. The equation for the total number density as a function of the chemical potential, $n(\mu) = -(1/\mathcal V) \partial\Omega(\mu)/\partial \mu$, can then be rewritten in terms of the gas parameter $\gamma = a_s n^{1/3}$ and the adimensional chemical potential $\mu^* \equiv (m/4\pi\hbar^2 n^{2/3})\mu$ as
\begin{equation}
    \frac{\mu^*}{\gamma} - \frac{32}{3\sqrt\pi}({\mu^*})^{3/2} = 1.
\end{equation}
This has exactly one solution for $\gamma \in [0,\,\gamma_c]$, where $\gamma_c = (\pi/768)^{1/3} \simeq 0.1599$, and no solutions for $\gamma > \gamma_c$. This places an upper limit, based on thermodynamic stability, on the value of $\gamma$ up to which a Gaussian theory with zero-range potential is applicable.

One possible solution to this problem is the inclusion of an effective finite-range interaction \cite{cappellaro}. Maintaining the form of the zero-range potential, one should instead consider quantum corrections beyond the Gaussian level. These corrections come from the terms in the action that are higher than second-order in the fluctuations $\psi_1$, $\psi_1^*$, given by \cite{stoof}
\begin{equation}
    S_\text{int}[\psi_1^*, \psi_1] = g \int_0^{\hbar\beta}d\tau\int d^3\mathbf x\left[\psi_0^* \psi_1(\mathbf x,\tau)|\psi_1(\mathbf x,\tau)|^2 + \psi_0 \psi_1^*(\mathbf x,\tau)|\psi_1(\mathbf x,\tau)|^2 +\frac{1}{2}|\psi_1(\mathbf x,\tau)|^4\right].
\end{equation}
In a perturbative expansion, these interaction terms generate Feynman diagrams contributing to the self-energy of $\psi_1$, which consists of all possible repetitions of the proper self-energy $\Sigma$ given by the sum of all irreducible diagrams. The repetitions of $\Sigma$ are resummed through the Dyson equation \cite{fetter}, giving for the corrected Green's function in momentum space
\begin{equation}
    \mathcal G(\mathbf k, \omega_n) = \frac{{\mathcal G}_\text{B}(\mathbf k, \omega_n)}{1-{\mathcal G}_\text{B}(\mathbf k, \omega_n)\Sigma(\mathbf k, \omega_n)},\label{dyson}
\end{equation}
where ${\mathcal G}_\text{B}$ is the Guassian-level Green's function,
\begin{align}
    -\mathcal G_\text{B}(\mathbf k, \omega_n) = \bigl\langle \psi_1(\mathbf k, \omega_n) \psi_1^*(\mathbf k, \omega_n)\bigr\rangle_\text{B} = \frac{\hbar(C_k + i\hbar \omega_n)}{(\hbar\omega_n)^2 + E_k^2(n_0)}.\label{gbog}
\end{align}
Here $\omega_n$ are bosonic Matsubara frequencies \cite{stoof,fetter}, $C_k \equiv \varepsilon_k + gn_0$, and $E_k(n_0)$ is the Bogoliubov spectrum. Using \eqref{gbog} and \eqref{dyson}, the density of noncondensed particles is then equal to
\begin{equation}
    n-n_0 = -\frac{1}{\hbar\beta \mathcal V}\sum_{\omega_n}\sum_{\mathbf k\neq \mathbf 0}\mathcal G(\mathbf k, \omega_n) = \frac{1}{\beta\mathcal V}\sum_{\omega_n}\sum_{\mathbf k\neq \mathbf 0}\frac{C_k + i\hbar\omega_n}{(\hbar\omega_n)^2 + E_k^2(n_0) + \hbar\Sigma(\mathbf k, \omega_n)\left(C_k + i\hbar\omega_n\right)}
\end{equation}
The zero-temperature quantum corrections to the Bogoliubov theory were addressed by Braaten and Nieto\cite{braaten}. Generalization of these results to finite temperature is an interesting prospect for future work.

In conclusion, in this paper we have discussed the effects of weak repulsive interactions on the condensation of a homogeneous Bose gas in 3D. Explicit equations for the condensate and superfluid fraction at finite temperature have been obtained within a variationally-optimized one-loop approximation. We have shown that at finite temperature, a small repulsive interaction reinforces the quantum statistical tendency to undergo BEC, so that as the interaction is switched on a sudden increase in the condensate fraction is
predicted. For larger values of gas parameter, instead, the condensate fraction decreases, as the stronger tendency to condense is overcome by the scattering of particles out of the condensate. Therefore, differently from the superfluid fraction, which at finite temperature is always increasing with the interaction strength, the condensate fraction is a non-monotonic function, presenting a global maximum for a small non-zero value of the gas parameter, which increases with the temperature. This prediction is robust, since the same qualitative behavior is observed using different approaches, although as we have discussed, the self-consistent Popov approximation is the most accurate in the region of interest. The semiclassical treatment of finite size effects shows that in a finite volume the non-monotonic behavior is even more evident.

Furthermore, repulsive interactions induce a positive shift of the critical temperature for BEC with respect to the noninteracting case, which in the self-consistent Popov approximation is predicted to be proportional to the square root of the gas parameter. This behavior is not captured by the standard Bogoliubov theory nor the Hartree-Fock-Bogoliubov approximation considered by Yuakalov and Yuakalova\cite{yukalov}. Combined with the quantum depletion at zero temperature, the positive shift of $T_c$ makes it clear why the condensed fraction must be non-monotonic in the gas parameter. 

This non-monotonic behavior has not yet been observed experimentally. In fact, based on our theoretical results we argue that experiments conducted so far have not concentrated in the range of interaction strength and temperature suitable for verifying the phenomenon. We believe however that this should be possible with the available experimental setups of ultracold atoms confined in a box potential. The measured condensate fraction will highlight the reliability of the analytical methods we have presented. 

\section{Methods}

\subsection*{Derivation of the condensate fraction from the grand canonical potential}
Eq. \eqref{pot} is the most general expression for the grand canonical potential of our system, because $\mu$ and $n_0$ have been kept as independent variables. This allows us to derive an equation for
$n$ as a function of $n_0$ and $T$, from which the condensate fraction follows. In particular,
\beqa\label{den}
  n(n_0,T) =-\frac{1}{\mathcal V}\frac{\partial\Omega(\mu, n_0, T)}
  {\partial \mu}\biggr|_{\mu = g n_0}= -\frac{1}{\mathcal V}\frac{\partial \bigl(\Omega_0 + \Omega_g^{(0)} + \Omega_g^{(T)}\bigr)}
  {\partial \mu}\biggr|_{\mu = g n_0} = n_0 + n_g^{(0)} + n_g^{(T)},
\eeqa
where the thermal density is
\begin{equation}
  n_g^{(T)} = \int \frac{d^3\mathbf k}{(2\pi)^3}\frac{\varepsilon_k
    + gn_0}{\left(\mathrm e^{\beta E_k(n_0)}-1\right)
    E_k(n_0)},
\end{equation}
while the zero-temperature density is
\begin{align}
  n_g^{(0)} &= \frac{1}{2}\int \frac{d^3\mathbf k}{(2\pi)^3}\frac{\varepsilon_k + gn_0}{E_k(n_0)}\biggr|_\text{UV reg.} = \frac{1}{3\pi^2}\left(\frac{m}{\hbar^2}g n_0\right)^{3/2}.
\end{align}
The second equality follows from the regularization of the integral, which is divergent in the ultraviolet, either by introducing a UV cutoff or by dimensional regularization\cite{salasnich}.

\subsection*{Analytical approximations for small temperature or small gas parameter}

Introducing the variable
$x = \beta \varepsilon_k$, the thermal contribution in Eq. \eqref{n} becomes
\begin{equation}\label{ngt2}
      n_g^{(T)} = \frac{\beta}{2\pi^2}\int_0^\infty dx\,\frac{dk(x)}{dx}
      \frac{k^2(x)}{(e^x-1)x}\left(\frac{\hbar^2k^2(x)}{2m}+gn_0\right)
\end{equation}
where 
\begin{equation}\label{k(x)}
      k(x) = \sqrt{\frac{2mgn_0}{\hbar^2}}\sqrt{-1+\sqrt{1+
          \frac{(k_\text{B}T)^2x^2}{g^2n_0^2}}}.
\end{equation}
In the limit $T\to 0$, $k(x)\simeq \sqrt{m/g n_0}(k_\text B T/\hbar)x$. The same expression for $k(x)$ may obtained by approximating the Bogoliubov spectrum with its phononic behavior, $E_k(n_0) \simeq c_s\hbar k$, i.e. the low temperature limit is equivalent to considering just phononic excitations. In this limit Eq. \eqref{ngt2} is calculated analytically and yields Eq. \eqref{smallt}.

In the limit $g \to 0$, $E_k(n_0) \simeq \hbar^2k^2/2m + gn_0$ and the integral on the right-hand side of Eq. (\ref{n}) is calculated analytically, giving Eq. \eqref{lowg}. {\cblue At finite temperature}, for $g\to 0$ the condensate fraction tends to the finite value expected for the ideal gas with infinite first derivative. In fact, from Eq. \eqref{lowg} we get
\begin{equation}\label{der1}
\frac{d}{dg}\frac{n_0}{n} = -\frac{n_0}{n^2}
\left[c_1\frac{d}{dg}\text{Li}_{3/2}\left(\mathrm e^{-\beta g n_0}\right)
  + c_2\sqrt g\right]
\end{equation}
with $c_{1}$, $c_2$ independent of $g$. Since $\frac{d}{dg}\text{Li}_{3/2}\left(\mathrm e^{-\beta g n_0}\right)
= -\beta n_0\, \text{Li}_{1/2}\left(\mathrm e^{-\beta g n_0}\right) \to -\infty$ for $g\to 0$,
\begin{equation}\label{der2}
\lim_{g\to 0} \frac{d}{dg}\frac{n_0}{n} = +\infty.
\end{equation}
{\cblue The first derivative of $n_0/n$ is thus positive (for $g \to 0$) due to the presence of the factor $\text{Li}_{3/2}\left(\mathrm e^{-\beta g n_0}\right)$, which is related to the thermal density. At zero temperature this factor is absent, and the condensate fraction decreases monotonically with $g$, as described by Eq. \eqref{condbog}.}

\section*{Acknowledgements}
The authors thank A. Pelster and R. P. Smith for useful discussions.
L. S. is partially supported by the European Union-NextGenerationEU within the National Center for HPC, Big Data and Quantum Computing 
[Project No. CN00000013, CN1 Spoke 10: "Quantum Computing"], 
by the BIRD Project "Ultracold atoms in curved geometries" of the 
University of Padova, by “Iniziativa Specifica Quantum” of Istituto Nazionale di Fisica Nucleare, by the European Quantum Flagship Project "PASQuanS 2", and by the PRIN 2022 Project "Quantum Atomic Mixtures: Droplets, Topological Structures, and Vortices" of the Italian Ministry for University and Research (MUR). Administrative and logistic support by the MUR project "Dipartimenti di Eccellenza: Frontiere Quantistiche" is also acknowledged.

\section*{Data availability statement}

Data are available upon request to L. Salasnich. 

\section*{ORCID IDs}
C Vianello https://orcid.org/0009-0001-1136-5924\\
L Salasnich https://orcid.org/0000-0003-0817-4753


\begin{thebibliography}{99}

\urlstyle{rm}
\expandafter\ifx\csname url\endcsname\relax
  \def\url#1{\texttt{#1}}\fi
\expandafter\ifx\csname urlprefix\endcsname\relax\def\urlprefix{URL }\fi
\expandafter\ifx\csname doiprefix\endcsname\relax\def\doiprefix{DOI: }\fi
\providecommand{\bibinfo}[2]{#2}
\providecommand{\eprint}[2][]{\url{#2}}

\bibitem{einstein}
\bibinfo{author}{Einstein, A.}
\newblock \bibinfo{journal}{\bibinfo{title}{Quantentheorie des einatomigen idealen gases}}.
\newblock {\emph{\JournalTitle{Sitzungsber. Preuss. Akad. Wiss., Phys. Math. Kl.
  Bericht}}} \textbf{\bibinfo{volume}{1}},
  \bibinfo{pages}{3} (\bibinfo{year}{1925}).

\bibitem{bose}
\bibinfo{author}{Bose, S. N.}
\newblock \bibinfo{journal}{\bibinfo{title}{Plancks Gesetz und Lichtquantenhypothese}}.
\newblock {\emph{\JournalTitle{Z. Phys.}}} \textbf{\bibinfo{volume}{26}},
  \bibinfo{pages}{178} (\bibinfo{year}{1924}).

\bibitem{penrose}
\bibinfo{author}{Penrose, O.} \& \bibinfo{author}{Onsager, L.}
\newblock \bibinfo{journal}{\bibinfo{title}{Bose-Einstein Condensation and Liquid Helium}}.
\newblock {\emph{\JournalTitle{Phys. Rev.}}} \textbf{\bibinfo{volume}{104}},
  \bibinfo{pages}{576} (\bibinfo{year}{1956}).

\bibitem{leggetts}
\bibinfo{author}{Leggett, A. J.}
\newblock \bibinfo{journal}{\bibinfo{title}{Superfluidity}}.
\newblock {\emph{\JournalTitle{Rev. Mod. Phys.}}} \textbf{\bibinfo{volume}{71}},
  \bibinfo{pages}{318} (\bibinfo{year}{1999}).

\bibitem{leggett}
\bibinfo{author}{Leggett, A. J.}
\newblock \bibinfo{journal}{\bibinfo{title}{Bose-Einstein condensation in the alkali gases: Some fundamental concepts}}.
\newblock {\emph{\JournalTitle{Rev. Mod. Phys.}}} \textbf{\bibinfo{volume}{73}},
  \bibinfo{pages}{307} (\bibinfo{year}{2001}).

\bibitem{andersen}
\bibinfo{author}{Andersen, J. O.}
\newblock \bibinfo{journal}{\bibinfo{title}{Theory of the weakly interacting Bose gas}}.
\newblock {\emph{\JournalTitle{Rev. Mod. Phys.}}} \textbf{\bibinfo{volume}{76}},
  \bibinfo{pages}{599} (\bibinfo{year}{2004}).

\bibitem{yukalov}
\bibinfo{author}{Yukalov, V. I.} \& \bibinfo{author}{Yukalova, E. P.}
\newblock \bibinfo{journal}{\bibinfo{title}{Condensate and superfluid fractions for varying interactions and temperature}}.
\newblock {\emph{\JournalTitle{Phys. Rev. A}}} \textbf{\bibinfo{volume}{76}},
  \bibinfo{pages}{013602} (\bibinfo{year}{2007}).

\bibitem{yukalov2}
 \bibinfo{author}{Yukalov, V. I.} \& \bibinfo{author}{Kleinert, H.}
\newblock \bibinfo{journal}{\bibinfo{title}{Gapless Hartree-Fock-Bogoliubov approximation for Bose gases}}.
\newblock {\emph{\JournalTitle{Phys. Rev. A}}} \textbf{\bibinfo{volume}{73}},
  \bibinfo{pages}{063612} (\bibinfo{year}{2006}).

\bibitem{yy2}
\bibinfo{author}{Yukalov, V.I.}\& \bibinfo{author}{Yukalova, E.P.}
\newblock \bibinfo{journal}{\bibinfo{title}{Bose-Einstein-condensed gases with arbitrary strong interactions}}.
\newblock {\emph{\JournalTitle{Phys. Rev. A}}} \textbf{\bibinfo{volume}{74}},
  \bibinfo{pages}{063623} (\bibinfo{year}{2006}).

\bibitem{cornell}
\bibinfo{author}{Anderson, M. H.}, \bibinfo{author}{Ensher, J. R.},
  \bibinfo{author}{Matthews, M. R.},
  \bibinfo{author}{Wieman, C. E.} \&
  \bibinfo{author}{Cornell, E. A.}
\newblock \bibinfo{journal}{\bibinfo{title}{Observation of bose-einstein condensation in a dilute atomic vapor}}.
\newblock {\emph{\JournalTitle{Science}}} \textbf{\bibinfo{volume}{269}},
  \bibinfo{pages}{198--201} (\bibinfo{year}{1995}).

\bibitem{ketterle}
\bibinfo{author}{Davis, K. B.},
\bibinfo{author}{Mewes, M.-O.},
\bibinfo{author}{Andrews, M. R.},
\bibinfo{author}{van Druten, N. J.},
\bibinfo{author}{Durfee, D. S.},
\bibinfo{author}{Kurn, D. M.} \&
\bibinfo{author}{Ketterle, W.}
\newblock \bibinfo{journal}{\bibinfo{title}{Bose-Einstein Condensation in a Gas of Sodium Atoms}}.
\newblock {\emph{\JournalTitle{Phys. Rev. Lett.}}} \textbf{\bibinfo{volume}{75}},
  \bibinfo{pages}{3969} (\bibinfo{year}{1995}).

\bibitem{nonuniform1}
\bibinfo{author}{Ho, T.-L.} \& \bibinfo{author}{Zhou, Q.}
\newblock \bibinfo{journal}{\bibinfo{title}{Obtaining the phase diagram and thermodynamic quantities of bulk systems from the densities of trapped gases}}.
\newblock {\emph{\JournalTitle{Nat. Phys.}}} \textbf{\bibinfo{volume}{6}},
  \bibinfo{pages}{131--134} (\bibinfo{year}{2010}).

\bibitem{nonuniform2}
\bibinfo{author}{Smith, R. P.},
\bibinfo{author}{Tammuz, N.},
\bibinfo{author}{Campbell, R. L. D.},
\bibinfo{author}{Holzmann, M.} \&
\bibinfo{author}{Hadzibabic, Z.}
\newblock \bibinfo{journal}{\bibinfo{title}{Condensed Fraction of an Atomic Bose Gas Induced by Critical Correlations}}.
\newblock {\emph{\JournalTitle{Phys. Rev. Lett.}}} \textbf{\bibinfo{volume}{107}},
  \bibinfo{pages}{190403} (\bibinfo{year}{2011}).

\bibitem{nonuniform3}
\bibinfo{author}{Qu, C.},
  \bibinfo{author}{Pitaevskii, L. P.} \&
  \bibinfo{author}{Stringari, S.}
\newblock \bibinfo{journal}{\bibinfo{title}{Expansion of harmonically trapped interacting particles and time dependence of the contact}}.
\newblock {\emph{\JournalTitle{Phys. Rev. A}}} \textbf{\bibinfo{volume}{94}},
  \bibinfo{pages}{063635} (\bibinfo{year}{2016}).

\bibitem{box-experiment1} 
\bibinfo{author}{Gaunt, A. L.},
\bibinfo{author}{Schmidutz, T. F.},
\bibinfo{author}{Gotlibovych, I.},
\bibinfo{author}{Smith, R. P.} \&
\bibinfo{author}{Hadzibabic, Z.}
\newblock \bibinfo{journal}{\bibinfo{title}{Bose-Einstein Condensation of Atoms in a Uniform Potential}}.
\newblock {\emph{\JournalTitle{Phys. Rev. Lett.}}} \textbf{\bibinfo{volume}{110}},
  \bibinfo{pages}{200406} (\bibinfo{year}{2013}).

\bibitem{smith}
\bibinfo{author}{Lopes, R.},
\bibinfo{author}{Eigen, C.},
\bibinfo{author}{Navon, N.},
\bibinfo{author}{Clement, D.},
\bibinfo{author}{Smith, R. P} \&
\bibinfo{author}{Hadzibabic, Z.}
\newblock \bibinfo{journal}{\bibinfo{title}{Quantum Depletion of a Homogeneous Bose-Einstein Condensate}}.
\newblock {\emph{\JournalTitle{Phys. Rev. Lett.}}} \textbf{\bibinfo{volume}{119}},
  \bibinfo{pages}{190404} (\bibinfo{year}{2017}).

\bibitem{box-experiment2} 
\bibinfo{author}{Ville, J. L.},
\bibinfo{author}{Saint-Jalm, R.},
\bibinfo{author}{Le Cerf, E.},
\bibinfo{author}{Aidelsburger, M.},
\bibinfo{author}{Nascimbene, S.},
\bibinfo{author}{Dalibard, J.} \&
\bibinfo{author}{Beugnon, J.}
\newblock \bibinfo{journal}{\bibinfo{title}{Sound Propagation in a Uniform Superfluid Two-Dimensional Bose Gas}}.
\newblock {\emph{\JournalTitle{Phys. Rev. Lett.}}} \textbf{\bibinfo{volume}{121}},
  \bibinfo{pages}{145301} (\bibinfo{year}{2018}).
      
\bibitem{box-experiment3} 
\bibinfo{author}{Navon, N.},
\bibinfo{author}{Smith, R. P} \&
\bibinfo{author}{Hadzibabic, Z.}
\newblock \bibinfo{journal}{\bibinfo{title}{Quantum gases in optical boxes}}.
\newblock {\emph{\JournalTitle{Nat. Phys.}}} \textbf{\bibinfo{volume}{17}},
  \bibinfo{pages}{1334--1341} (\bibinfo{year}{2021}).

\bibitem{salasnich}
\bibinfo{author}{Salasnich, L.} \& \bibinfo{author}{Toigo, F.}
\newblock \bibinfo{journal}{\bibinfo{title}{Zero-point energy of ultracold atoms}}.
\newblock {\emph{\JournalTitle{Phys. Rep.}}} \textbf{\bibinfo{volume}{640}},
  \bibinfo{pages}{1} (\bibinfo{year}{2016}).

\bibitem{stoof}
\bibinfo{author}{Stoof, H. T. C.}, 
\bibinfo{author}{Gubbels, K. B.} \&
\bibinfo{author}{Dickerscheid, D. B. M.}
\newblock \emph{\bibinfo{title}{Ultracold Quantum Fields}}.
(\bibinfo{publisher}{Springer}, \bibinfo{year}{2009}).

\bibitem{bogoliubov1}
\bibinfo{author}{Bogoliubov, N. N.}
\newblock \emph{\bibinfo{title}{Lectures on Quantum Statistics, Vol. 1}}.
(\bibinfo{publisher}{Gordon and Breach}, \bibinfo{year}{1967}).

\bibitem{bogoliubov2}
\bibinfo{author}{Bogoliubov, N. N.}
\newblock \emph{\bibinfo{title}{Lectures on Quantum Statistics, Vol. 2}}.
(\bibinfo{publisher}{Gordon and Breach}, \bibinfo{year}{1967}).

\bibitem{bogo_orig}
\bibinfo{author}{Bogoliubov, N. N.}
\newblock \bibinfo{journal}{\bibinfo{title}{On the Theory of Superfluidity}}.
\newblock {\emph{\JournalTitle{J. Phys. (USSR)}}} \textbf{\bibinfo{volume}{11}},
  \bibinfo{pages}{23} (\bibinfo{year}{1947}).

\bibitem{kleinert} 
\bibinfo{author}{Kleinert, H.},
\bibinfo{author}{Schmidt, S.} \&
\bibinfo{author}{Pelster, A.} 
\newblock \bibinfo{journal}{\bibinfo{title}{Quantum phase diagram for homogeneous Bose-Einstein condensate}}.
\newblock {\emph{\JournalTitle{Ann. Phys. (Leipzig)}}} \textbf{\bibinfo{volume}{14}},
  \bibinfo{pages}{214} (\bibinfo{year}{2005}).

\bibitem{stevenson} 
\bibinfo{author}{Stevenson, P. M.}
\newblock \bibinfo{journal}{\bibinfo{title}{Optimized Perturbation Theory}}.
\newblock {\emph{\JournalTitle{Phys. Rev. D}}} \textbf{\bibinfo{volume}{23}},
  \bibinfo{pages}{2916} (\bibinfo{year}{1981}).

\bibitem{svistunov}
\bibinfo{author}{Kashurnikov, V. A.},
\bibinfo{author}{Prokof'ev, N. V.} \&
\bibinfo{author}{Svistunov, B. V.}
\newblock \bibinfo{journal}{\bibinfo{title}{Critical Temperature Shift in Weakly Interacting Bose Gas}}.
\newblock {\emph{\JournalTitle{Phys. Rev. Lett.}}} \textbf{\bibinfo{volume}{87}},
  \bibinfo{pages}{120402} (\bibinfo{year}{2001}).

\bibitem{arnold}
\bibinfo{author}{Arnold, P.} \&
\bibinfo{author}{Moore, G.}
\newblock \bibinfo{journal}{\bibinfo{title}{BEC Transition Temperature of a Dilute Homogeneous Imperfect Bose Gas}}.
\newblock {\emph{\JournalTitle{Phys. Rev. Lett.}}} \textbf{\bibinfo{volume}{87}},
  \bibinfo{pages}{120401} (\bibinfo{year}{2001}).

\bibitem{kleinert2}
\bibinfo{author}{Kleinert, H.}
\newblock \bibinfo{journal}{\bibinfo{title}{Five-Loop Critical Temperature Shift in Weakly Interacting Homogeneous Bose–Einstein Condensate}}.
\newblock {\emph{\JournalTitle{Mod. Phys. Lett. B}}} \textbf{\bibinfo{volume}{17}},
  \bibinfo{pages}{1011--1020} (\bibinfo{year}{2003}).

\bibitem{kastening}
\bibinfo{author}{Kastening, B.}
\newblock \bibinfo{journal}{\bibinfo{title}{Bose-Einstein condensation temperature of a homogenous weakly interacting Bose gas in variational perturbation theory through seven loops}}.
\newblock {\emph{\JournalTitle{Phys. Rev. A}}} \textbf{\bibinfo{volume}{69}},
  \bibinfo{pages}{043613} (\bibinfo{year}{2004}).

\bibitem{rossi}
\bibinfo{author}{Rossi, M.} \&
\bibinfo{author}{Salasnich, L.}
\newblock \bibinfo{journal}{\bibinfo{title}{Path-integral ground state and superfluid hydrodynamics of a bosonic gas of hard spheres}}.
\newblock {\emph{\JournalTitle{Phys. Rev. A}}} \textbf{\bibinfo{volume}{88}},
  \bibinfo{pages}{053617} (\bibinfo{year}{2013}).

\bibitem{landau}
\bibinfo{author}{Landau, L. D.} \&
\bibinfo{author}{Lifshitz, E. M.}
\newblock \emph{\bibinfo{title}{Course of Theoretical Physics, Vol. 9.
    Statistical Physics: Theory of the condensate State}}.
(\bibinfo{publisher}{Butterworth-Heinemann}, \bibinfo{year}{1980}).

\bibitem{lanaro}
\bibinfo{author}{Lanaro, M.},
\bibinfo{author}{Bighin, G.},
\bibinfo{author}{Dell'Anna, L.} \&
\bibinfo{author}{Salasnich, L.}
\newblock \bibinfo{journal}{\bibinfo{title}{Finite-size effects in the two-dimensional BCS-BEC crossover}}.
\newblock {\emph{\JournalTitle{Phys. Rev. B}}} \textbf{\bibinfo{volume}{109}},
  \bibinfo{pages}{104511} (\bibinfo{year}{2024}).

\bibitem{lhy}
\bibinfo{author}{Lee, D. T.},
\bibinfo{author}{Huang, K.} \&
\bibinfo{author}{Yang, C. N.}
\newblock \bibinfo{journal}{\bibinfo{title}{Eigenvalues and Eigenfunctions of a Bose System of Hard Spheres and Its Low-Temperature Properties}}.
\newblock {\emph{\JournalTitle{Phys. Rev.}}} \textbf{\bibinfo{volume}{106}},
  \bibinfo{pages}{1135} (\bibinfo{year}{1957}).

\bibitem{cappellaro}
\bibinfo{author}{Cappellaro, A.} \&
\bibinfo{author}{Salasnich, L.}
\newblock \bibinfo{journal}{\bibinfo{title}{Thermal field theory of bosonic gases with finite-range effective interaction}}.
\newblock {\emph{\JournalTitle{Phys. Rev. A}}} \textbf{\bibinfo{volume}{95}},
  \bibinfo{pages}{033627} (\bibinfo{year}{2017}).

\bibitem{fetter}
\bibinfo{author}{Fetter, A. L.} \&
\bibinfo{author}{Walecka, J. D.}
\newblock \emph{\bibinfo{title}{Quantum Theory of Many-Particle Systems}}.
(\bibinfo{publisher}{Dover Publications}, \bibinfo{year}{2003}).

\bibitem{braaten}
\bibinfo{author}{Braaten E.} \&
\bibinfo{author}{Nieto A.}
\newblock \bibinfo{journal}{\bibinfo{title}{Quantum corrections to the energy density of a homogeneous Bose gas}}.
\newblock {\emph{\JournalTitle{Eur. Phys. J. B}}} \textbf{\bibinfo{volume}{11}},
  \bibinfo{pages}{143--159} (\bibinfo{year}{1999}).
       
\end{thebibliography}
\end{document}